\documentstyle[12pt]{article}

\newcommand\be{\begin{equation}}
\newcommand\ee{\end{equation}}
\newcommand\bea{\begin{eqnarray}}
\newcommand\eea{\end{eqnarray}}

\newcommand\lsim{\mathrel{\rlap{\lower4pt\hbox{\hskip1pt$\sim$}}
	\raise1pt\hbox{$<$}}}
\newcommand\gsim{\mathrel{\rlap{\lower4pt\hbox{\hskip1pt$\sim$}}
	\raise1pt\hbox{$>$}}}

\newcommand\mpl{M_{\rm Pl}}

\begin{document}

\title{Non-abelian discrete gauge symmetries and inflation}

\bigskip
\author{
J. D. Cohn\footnote{jcohn@cfa.harvard.edu} \\
{\normalsize \em
 Depts. of Physics and Astronomy, U. Illinois,
 Urbana, IL 61801 and} \\
{\normalsize \em 
 Harvard-Smithsonian Center for 
Astrophysics \footnote{Address from August 1999.} }\\
{\normalsize \em  60 Garden St., Cambridge, MA 02138
}\\
{}\\
E. D. Stewart\footnote{ewan@kaist.ac.kr} \\
{\normalsize \em
 NASA/Fermilab Astrophysics Group,
 FNAL, Batavia, IL 60510-0500 and }\\
{\normalsize \em
 Department of Physics
KAIST, Taejon 305-701, Korea\footnote{Address from August 1999.}
}
{}\\
{}\\
{\normalsize KAIST-TH 00/2}
}
\maketitle
\begin{abstract}

{\normalsize
Obtaining a potential flat enough to provide slow roll inflation is
often difficult when gravitational effects are included.
Non-abelian
discrete gauge symmetries can guarantee the flatness of the inflaton
potential in this case, and also provide special field values where
inflation can end.}
\end{abstract}
\setcounter{page}{0}
\newpage
\setcounter{page}{1}

\section{Introduction}

Inflation provides initial conditions in the early universe with
attractive and observationally motivated features
\cite{inflation}.
When inflation is driven by the 
almost constant potential energy $V(\phi)$ of
a field $\phi$, the equation of motion for $\phi(t)$
($\phi$ becomes spatially homogeneous quickly after inflation starts) is
\be
\ddot{\phi} + 3 H \dot{\phi} = -\frac{\partial V(\phi)}{\partial \phi}
\ee
where 
\be
(\frac{\dot{a}}{a})^2 =
H^2 = \frac{8 \pi G}{3} \rho .
\ee
Here
$\rho$ is the energy density, $a(t)$ the
scale factor, and the curvature scale and cosmological constant
have been taken to be zero.

Approximately scale invariant perturbations are consistent with
observation and can be produced with slow roll
inflation\cite{slowroll}.
Slow roll inflation not only requires the inflationary
acceleration of the scale factor ($\ddot{a} > 0$ which here
becomes
$V \gg \dot{\phi}^2$) but also that
\be
\label{vpcond}
\left( \frac{V'}{V} \right)^2 \ll \frac{1}{\mpl^2}
\ee
and
\be
\label{vppcond}
\left| \frac{V''}{V} \right| \ll \frac{1}{\mpl^2} \; .
\ee
Here $\mpl$ is the Planck constant, $1/\sqrt{8 \pi G}$. 
Many models of inflation are built ignoring gravitational
strength interactions, and so are implicitly setting
$ M_{\rm Pl} = \infty $.  In this case, satisfying the
 conditions above
is not possible.
Keeping $M_{\rm Pl}$ finite implies one should include gravitational
effects.  To include gravitational effects,
we will work in the context of supergravity.  In this case,
the first condition suggests we should be near a maximum, or other
extremum, of the potential.
The second has been found to be non-trivial \cite{fvi,iss}.
In supergravity, the potential is composed of two parts, the $F$-term and
the $D$-term.
If the inflationary potential energy is dominated by the $F$-term then one
can show that \cite{Dine,fvi,iss}
\be
\frac{V''}{V} = \frac{1}{\mpl^2} + \mbox{model dependent terms}
\ee
Unless the model dependent terms cancel the first term, the second slow
roll condition, Eq.~(\ref{vppcond}) above, is violated.
Thus to build a model of slow-roll inflation one must be able to control 
the gravitational strength corrections.
There have been various attempts at achieving slow-roll inflation 
naturally, for a discussion see \cite{nat-infl,longer},
for a recent review of inflationary models see, e.g. \cite{ly-rio}.

Discrete non-abelian gauge symmetries provide several ingredients
for successful inflation.  The symmetries ensure that the potential is 
flat enough for inflation to occur, and the discreteness of the symmetries
provide exit points in the potential where inflation can end.
As will be seen below, non-abelian symmetries in particular
are useful for hybrid models in low energy effective field
theories with supergravity corrections.

\section{The idea}\label{idea}
In hybrid\cite{hybrid} and mutated\cite{mutated} 
hybrid models, one field slowly rolls
and a second is pinned until the first field reaches some
critical value.  E.g., for concreteness and illustration,
take the slowly rolling field to be
$\Phi$ and call the second field $\Psi$.
The hybrid potential for $\Psi$ in the simplest case has the form
\be
\label{hybridpot}
(\alpha |\Phi|^\beta - m^2) |\Psi|^2 + O(\Psi^3)
\ee
where $m^2, \; \alpha, \; \beta > 0$.  The potential for $\Phi$ is
not shown, its crucial feature is sufficient flatness to allow for
slow roll (the conditions (\ref{vpcond},\ref{vppcond})).  
For large $\Phi$, the field $\Psi$ has a large mass and is pinned.  
During slow roll inflation, $\Phi$ decreases, until near the 
critical value $\Phi_c^\beta \sim m^2/\alpha$ the field $\Psi$ is freed.
The rolling of $\Psi$ eventually ends inflation, often
quickly after $\Phi$ reaches the critical value.  The specific 
properties of the exit depend on the other terms in the potential
as well.
(Slow roll conditions for multiple fields are discussed in \cite{sr-mult}).
Supergravity corrections, as mentioned above, will generically
give both $\Phi$ and $\Psi$ masses of order the Hubble 
constant, making slow roll for $\Phi$ 
hard to achieve.

This letter describes a mechanism for providing a flat potential
when supergravity corrections are included, and providing a way for
inflation to end via special exit points.  Although hybrid 
models are discussed here, a variant also works for mutated 
hybrid models as well.  Hybrid and mutated hybrid 
inflationary models have the advantage that the energy scale of 
inflation can sometimes be quite low, which  allows for a description in 
terms of lower energy, and thus often better controlled, theories.
(The low energy scale advantage is coupled with the
caveat that large fluctuations are likely at the end of inflation in
many of these models\cite{spike}.)

Adding a discrete non-abelian gauged symmetry in the low
energy effective field theory, the context for these models, 
provides for both the flatness and the exit as follows. 
Instead of taking one field $\Phi$ and one field $\Psi$, take
an N-tuple of fields, $\Phi_i$ and $\Psi_i$, with $i = 1, ...N$,
and impose a discrete non-abelian gauged symmetry.  This symmetry
has to be chosen such that the lowest order allowed terms
in the superpotential respect a larger continuous symmetry, for
example SU(N) in this case.  The higher order terms then break
this effective continuous symmetry to a discrete subgroup.
The other condition on this effective field theory is that
the supersymmetry breaking comes from some other sector, and
provides a vacuum energy $V_0$.  As we are using an effective field
theory we do not need to say where this term comes from in particular.
The vacuum energy in turn generates supersymmetry breaking
masses for the fields $\Phi_i, \Psi_i$ as well as other higher order terms.

As the lowest order terms in the superpotential respect the
SU(N) symmetry, the mass terms for the fields will be functions of
the SU(N) invariants $|\Phi|^2 = \sum_i |\Phi_i|^2$ and
$|\Psi|^2 = \sum_i|\Psi_i|^2$.  Renormalization group flow\cite{Witten} will
shift the masses $m_\Phi^2$ and $m_\Psi^2$ to have non-trivial
minima which to leading order only depend on the SU(N) symmetric
field combination $|\Phi|$ and $|\Psi|$, that is, for example,
\be
V(|\Phi_i|) \sim m_\Phi^2(|\Phi|-|\Phi_0|)^2  + \dots\; , \;
\ee
This consequently provides for flat directions in the potential where
inflation can occur.  The field $|\Phi|$ will be locked,
$|\Phi | \sim |\Phi_0|$ but the various components of $\Phi$ can vary, 
causing slow roll inflation.

If the SU(N) symmetry were exact, then the corresponding hybrid
term above would have to also respect the SU(N) symmetry and
be a function of $|\Phi|, |\Psi|$ alone, just as above:
\be
V(\Phi_i, \Psi_i) \sim (\alpha |\Phi|^\beta - m_\Psi^2) |\Psi|^2 + ...
\ee
Although the components of $|\Phi|$, the $\Phi_i$, are
free to vary, only their combination appears in
the hybrid exit, and as
$|\Phi|$ remains locked, the exit cannot take place.
On the other hand, if the theory only respects a discrete
subgroup of SU(N) this can change.
For instance if the theory respects say a permutation symmetry,
the hybrid term 
\be
V(\Phi_i,\Psi_i) \sim 
\sum_i (\alpha_i |\Phi_i|^\beta - m_\Psi^2) |\Psi_i|^2 + h.o.t. 
\ee
may be allowed.
If $\beta \ge 4$, this term will be higher order than the couplings 
which generate the SU(N) symmetric masses, thus allowing the
SU(N) symmetry to consistently pin
$|\Phi| \sim |\Phi_0|$.
In this case, the individual $|\Phi_i|$ can control the end of
inflation, by decreasing to some critical value,
although the sum of the $|\Phi_i|^2$ is fixed by
the supersymmetry breaking and SU(N) respecting renormalization
group effects.  In this way,
the discrete symmetry allows inflation both to occur (via
the continuous effective symmetry) and to end with a hybrid mechanism
(via the discrete subgroup).  As the norms of the fields, the
$|\Phi_i|$ appear in these hybrid potentials, a non-abelian discrete
symmetry is needed.  
An abelian discrete gauge theory would allow for changes in the phase of 
$\Phi_i$, but as the exit depends on the magnitude of $\Phi_i$ 
this abelian variation will not trigger the exit.
The basic idea of using discrete gauged symmetries 
generalizes the idea used in Natural 
Inflation \cite{natural}.  
It is also similar to \cite{Ross}, where an approximate symmetry
is provided by lower terms in the lagrangian (in particular the kinetic
terms).  Although orbifold constructions, such as \cite{fvi,iss,orbifold}
also involve a discrete gauge symmetry, the flatness of the
potential is obtained differently.

\section{Example}

This basic idea can be used to construct a low energy effective
model.
One chooses a suitable gauge group and representations, ensuring that
the gauge symmetries are not anomalous.  The
symmetries constrain the allowed terms in the superpotential $W$ and
Kahler potential $K$.  The couplings and supersymmetry
breaking terms correspond to regions in parameter space which 
have constraints both from observation and underlying physics
(for example the supersymmetry breaking terms will naturally have
coefficients the scale of supersymmetry breaking).
For this effective field theory, as mentioned
earlier, we assume a hidden sector breaks supersymmetry.
This generates supersymmetry breaking terms in the
effective potential, including a vacuum energy 
$V_0$ and masses for the scalars.  In addition, as also mentioned
earlier, the
renormalization group running of the supersymmetry
breaking mass term for $\Phi$ to generates a potential for $\Phi$ with non-
trivial minimum $|\Phi| = \Phi_0$.
The renormalization is induced (to leading order) by low dimension 
couplings symmetric under the extended continuous symmetry.
Thus the renormalization group masses and the potential will be symmetric
under the extended continuous symmetry.

A superpotential with a term $\Phi^{\frac{\beta}{2}} \Psi^2$,
will produce a
hybrid potential of the form above, eqn.(\ref{hybridpot}),
using the same notation for the scalar component of the superfield
and the superfield itself.  For example, a model based on
a discrete subgroup of SU(2) can have the fields and charges
shown 
in table one.
\begin{table}[h]
\begin{center}
\begin{tabular}{|c|cccc|}
\hline
& $\phi_i$ & $\psi_i$ & $\chi_i$ & $\rho$ \\
\hline
$SU(2)_4 \subset$ SU(2)
& $\bf{2}$ & $\bf{2}$ & $\bf{2}$ & $1$ \\
${\bf Z}_4$
& $1$ & $1$ & $-2$ & $1$ \\
${\bf Z}_M$
& $1$ & $-1$ & $0$ & $-1$ \\
\hline
\end{tabular}
\end{center}
\caption{\label{hybridtable}
Symmetries and fields.}
\end{table}
The second discrete symmetry must have
$M>4$, the specific value is not important.
The generators of the discrete subgroup of SU(2) are taken to
be
\be
\left(
\begin{array}{ll}
0&i\\
i&0 
\end{array} \right) \; \;
\left(
\begin{array}{ll}
0&1\\
-1&0 
\end{array} \right) \; \;
\left(
\begin{array}{ll}
e^{\frac{2 \pi i}{N}}&0\\
0& e^{-\frac{2 \pi i}{N}}
\end{array} \right) \; \;
\ee
with $N>4$.
The holomorphic terms which respect the symmetries
give the superpotential for this model:
\be
W = \lambda_0 (\phi_1 \chi_2 - \phi_2 \chi_1) \rho +
\frac{1}{2} \lambda_1(\phi_1^2 \psi_2^2 + \phi_2^2 \psi_1^2)
+ \lambda_2 \phi_1 \phi_2 \psi_1 \psi_2  + 
\lambda_\chi \chi_1^2 \chi_2^2 + h.o.t. \; .
\ee
The lowest order terms respect the full SU(2), while the higher
order terms only have this symmetry for special values of
$\lambda_1, \lambda_2$.  The leading neglected higher terms are of the form
$\rho \phi \chi^3$, $\phi^2 \chi^2 \rho^2$ and
$\phi^2 \psi^2 \chi^2$.
Calculating the bosonic contribution to the potential (again using the
same notation for the superfields and their bosonic components) from this
superpotential gives
\be
\begin{array}{ll}
V_{susy} = &
|\lambda_0|^2 (\phi_1 \chi_2 - \phi_2 \chi_1|^2 +
|\phi_1 \rho + \frac{\lambda_\chi}{\lambda_0} \chi_2 \chi_1^2|^2 + 
|-\phi_2 \rho+ \frac{\lambda_\chi}{\lambda_0} \chi_2^2 \chi_1|^2) \\
&+
|\lambda_0 \chi_2 \rho + \lambda_1\phi_1 \psi_2^2
+ \lambda_2 \phi_2 \psi_1 \psi_2|^2 
+|-\lambda_0 \chi_1 \rho + \lambda_1\phi_2 \psi_1^2
+ \lambda_2 \phi_1 \psi_1 \psi_2|^2 \\
&+|\lambda_1 \phi_2^2 \psi_1
+ \lambda_2 \phi_1 \phi_2 \psi_2|^2 +
|\lambda_1 \phi_1^2 \psi_2
+ \lambda_2 \phi_1 \phi_2 \psi_1|^2  + 
h.o.t.\\
\end{array}
\ee
The supersymmetry breaking terms are
\be
\begin{array}{ll}
V_{break} &=
V_0 + m_\chi^2|\chi|^2 + \tilde{m}_\phi^2 |\phi|^2 
-m_\psi^2|\psi|^2 + m_\rho^2 |\rho|^2 +
\mu_0 (\phi_1 \chi_2 - \phi_2 \chi_1) \rho \\
&+
\frac{1}{2} \mu_1(\phi_1^2 \psi_2^2 + \phi_2^2 \psi_1^2)
+ \mu_2 \phi_1 \phi_2 \psi_1 \psi_2  +
h.c. +  h.o.t.
\end{array}
\ee
Note that for the hybrid exit the fields $\psi_i$ have a
negative mass squared.
As mentioned earlier, we take the supersymmetry breaking mass term for
$\phi$ ($|\phi|^2 = |\phi_1|^2 + |\phi_2|^2$) to be its renormalized form
\be
\tilde{m}^2|\phi|^2 = m_\phi^2 (|\phi|-|\phi_0|)^2 + h.o.t.
\ee
The mass term for $\psi$ can have a similar form, but as it
will be considered near the value $\psi \sim 0$, this effect
is not relevant.

An extremum of the combined potential
$V_{susy} + V_{break}$
occurs when
\be
\label{background}
\chi_i = \rho = \psi_i = 0 , \; |\phi| \sim \phi_0.
\ee
This is the background value around which slow roll inflation
is taken to occur.
The mass terms for $\psi_i$ are
\be
-m_\psi^2|\psi|^2 
+\frac{1}{2} \mu_1(\phi_1^2 \psi_2^2 + \phi_2^2 \psi_1^2)
+ \mu_2 \phi_1 \phi_2 \psi_1 \psi_2  +
h.c. 
+|\lambda_1 \phi_2^2 \psi_1
+ \lambda_2 \phi_1 \phi_2 \psi_2|^2 +
|\lambda_1 \phi_1^2 \psi_2
+ \lambda_2 \phi_1 \phi_2 \psi_1|^2 
\ee
Taking $\lambda_2 \ll \lambda_1$ so that the SU(2) is strongly
broken, and using that $\mu_i$ are small, one disentangles
that the leading terms in the mass of $\psi_i$ are
proportional to $|\lambda_1 \epsilon_{ij}\phi_j^2|^2$.  Thus as mentioned
earlier $|\phi|$ can remain fixed while
$|\phi_j|$ rolls, and thus inflation can end.

\section{Summary}
We have shown that
non-abelian discrete gauge symmetries can provide for sufficient
flatness for inflation to occur and special exit points where
inflation can end.  
The flatness comes from an approximate continuous symmetry induced
by the discrete symmetry, while the special exit points and dynamics
are due to the fact that only the discrete symmetry is exact and
unbroken.
This gives a new way for inflation to happen
even when masses from supergravity corrections are taken into account.
As a low energy effective field theory is used, one does not need
specific information about physics at higher energies, but only
the induced field content and symmetries at the scale of interest.

Integrating this mechanism into a more complete model
requires several steps, which are outside the scope of this
letter.  The field content must be anomaly free
which may require the addition of more fields.  
Other considerations will constrain the model's parameter space and 
background field values.  For instance, the stability of the potential 
around eq.(\ref{background}) may require specific values of $\phi_0$ and 
perhaps couplings.  The slow roll behavior of $\phi_i$ must lead $\phi_i$
to the hybrid exit, and will be controlled by higher order corrections 
in the potential such as those mentioned above, corrections to the Kahler 
potential and supersymmetric loop corrections. The exit from inflation
must not introduce large fluctuations in an observable regime,
as mentioned earlier.  Constraints from the measured COBE
normalization\cite{cobe} and the scale dependence of the fluctuations
(tilt) must also be imposed.
These requirements fix combinations of the parameters for
the model.  In a longer forthcoming paper \cite{longer}, we
present two specific models utilizing discrete non-abelian 
gauge symmetry as described here. 
We impose these constraints for a hybrid and
mutated hybrid model, 
including higher order corrections, and calculate detailed
properties of the parameter space, the exit from inflation
and features of the spectrum.

EDS was supported by the DOE and the NASA grant NAG 5-7092 at
Fermilab,  Grant No. 1999-2-111-002-5 from the
interdisciplinary Research Program of the KOSEF and
Brain Korea 21 Project. 
JDC was supported by NSF-PHY-9800978 and NSF-PHY-9896019.
For hospitality, EDS thanks M. White and R. Leigh at 
UIUC, JDC thanks the Aspen Center for Physics, 
and we both thank the Santa Fe 99 Workshop on Structure Formation
and Dark Matter.
JDC is grateful to Martin White for numerous discussions, and
J. Terning for information about supersymmetry.

\end{document}